# Varying Electronic Coupling at Graphene-Copper Interfaces Probed with Raman Spectroscopy

Jieun Choi[1], Seonghyun Koo,[1] Myeongin Song,[1] Daeyool Jung[2], Sungyool Choi[2], and Sunmin Ryu[1,3]*

[1]Department of Chemistry, Pohang University of Science and Technology (POSTECH), Pohang, Gyeongbuk 37673, Korea

[2]School of Electrical Engineering, Korea Advanced Institute of Science and Technology (KAIST), Daejeon 34141, Korea

[3]Division of Advanced Materials Science, Pohang University of Science and Technology (POSTECH), Pohang, Gyeongbuk 37673, Korea

* E-mail: sunryu@postech.ac.kr

As the synthesis of graphene on copper became one of the primary preparation methods for both fundamental research and industrial application, Raman spectra of graphene/Cu systems need to be quantitatively understood regarding how their interactions affect the electronic structure of graphene. Using multi-wavelength Raman spectroscopy, we investigated three types of graphene bound on Cu: graphene grown on Cu foils and Cu film/SiO$_2$, and Cu-evaporated exfoliated graphene. 2D peak frequencies of the first two samples were ~17 cm$^{-1}$ higher than expected for 1.96 eV excitation even when the effect of strain was considered. More notably, the upshift in 2D decreased with increasing excitation energy. Based on control experiments using Cu-evaporated graphene, we revealed that the spectral anomaly was induced by environment-dependent nonlinear dispersion in the electronic bands of graphene and determined the degree of the electronic modification. We also showed that the large upshifts of G and 2D peaks originating from differential thermal expansion of Cu could be significantly reduced by backing Cu films with dielectric substrates of insignificant thermal expansion. The quantitative analysis of electronic coupling between graphene and Cu presented in this study will be highly useful in characterizing as-grown graphene and possibly in other forms.

# 1. Introduction

Since graphene was first isolated onto dielectric substrates via a seemingly simple mechanical cleavage,[1] it has been a representative two-dimensional (2D) material at the focus of intensive research efforts across various academic disciplines.[2, 3] With high electrical mobility, optical transparency, and structural stability, graphene presents great potential in various applications ranging from electronics, optoelectronics, and transparent conducting electrodes to functional composite materials, coatings, and conductive inks.[4] Synthesis of graphene membranes with large size and good quality through chemical vapor deposition (CVD)[5, 6] has spurred industrial effort in converting the proposed ideas into realities. Because of its low carbon solubility, copper[5] as a CVD catalyst generates monolayer graphene selectively unlike nickel[7, 8] that leads to multilayer films. As the Cu-CVD method has been continuously improved in terms of crystallinity, domain size, and overall area,[9] Cu has distinct strengths over other catalysts in growing quality graphene. Remarkably, Xu et al. reported an oxygen-assisted fast CVD of meter-sized graphene single crystals using industry-grade Cu foils.[10] The Cu-graphene interface has also drawn research interests for long-term stability[11] and improvement of adhesion.[12] All of these recent developments are attracting considerable interest not only from the academic community but also from the related industry that awaits economically viable means for mass production.

Mass production and industrial application of graphene also require reliable and efficient characterization of its various physical and chemical properties.[13] In response to this, there have been intense efforts in developing various experimental probes that are optimized for graphene in various forms. Scanning tunneling microscopy is a real-space structural and electronic probe for graphene supported on conducting substrates.[14, 15] Transmission electron microscopy also enables one to see each carbon atom and visualize any structural irregularities in graphene that is standing free from supporting substrates.[16] Despite the extreme atomic spatial resolution, these two methods are inefficient in assessing large areas for statistical purpose and can be used only for samples that satisfy specific requirements for the measurements. Moreover, quality check and control in CVD production lines would need rapid on-site characterization of graphene that is still supported on metallic growth catalysts. In this regard, optical probes using scattering, reflection and ellipsometry[17] are suitable for this purpose as they can be operated with high throughput for large-area samples.[13] Although Raman spectroscopy has served as a powerful tool in characterizing thickness,[18-20] defects,[21-23] charge density,[24, 25] and lattice strain of graphene,[26, 27] however, its quantitative use has been limited for graphene supported on Cu substrates because of strong metallic photoluminescence[28] that interferes with Raman signals. Besides, the frequencies of G and 2D Raman peaks cannot be translated into charge density[21, 22] and lattice strain[26, 27] using the Raman metrology developed for graphene supported on dielectric substrates[29] because electronic interactions with metallic substrates may modify the effective Fermi velocity of graphene[30, 31] and thus affect the double resonance excitation of 2D peak.[32] Although such an effect manifested by unusually high 2D frequency has been observed for graphene grown on Cu foils and single crystals,[33, 34] its quantitative understanding has not been achieved. It also remains unknown how the effect depends on the excitation wavelength. These are essential questions not only because they will reveal the modified electronic structure of graphene but also multi-wavelength Raman spectroscopy needs to be established for efficient industrial graphene metrology.

In this work, we performed multi-wavelength Raman measurements for graphene in contact with Cu to determine electronic modification and its effect on Raman spectra over a wide range of energy. To investigate the effect of the degree of physical contact, we compared as-grown graphene on Cu with Cu-deposited graphene. Whereas G and 2D frequency are widely spread because of lattice strain that originates from differential thermal expansion between graphene and Cu, an additional significant upshift in 2D frequency was observed only for graphene in good contact with Cu and found to increase in magnitude with decreasing excitation photon energy. The Cu-induced hardening of 2D mode provides experimental maps of the electronic bands that are modified by the metallic substrates. This work completes the Raman metrology of graphene under the effects of strain, extra charges and electronic modification by environments.

## 2. Methods
**Preparation of CVD samples**. Two types of single-layer graphene samples were grown on Cu foils ($G_{CuFoil}$) and Cu thin films ($G_{CuFilm}$), respectively. Cu foils (Nilaco Corporation, 30 μm-thick, 99.9% purity) were dipped into 5% $HNO_3$ solution for 30 seconds to remove surface contamination before CVD growth. The residual acid solution was removed by thorough rinsing with deionized water followed by $N_2$ blow. As the catalyst for $G_{CuFilm}$, 300-nm Cu and 15-nm Ni films were sequentially deposited on Si wafers with 300-nm oxide layer by e-beam evaporation. Thin Ni films required to suppress evaporation of Cu films are known to diffuse into Cu film efficiently during CVD growth, and the top surfaces contain low concentration of Ni.[35] Indeed, X ray photoemission spectroscopy showed that the Ni/Cu atomic ratio in the top surface region of $G_{CuFilm}$ samples was 0.10 ~ 0.11, which was much lower than the value (~0.4) reported in the literature.[35] $G_{CuFoil}$ was grown with a methane flow of 8 mL/min in a tube furnace at 1000 °C for 60 min followed by pre-annealing in an $H_2$ flow at 1070 °C for reduction of copper oxide and smoother surface. The overall pressure was maintained at 500 mTorr during the growth. $G_{CuFilm}$ was synthesized in an ICP (inductively coupled plasma) CVD system. The Cu-deposited wafer was heated up to 960 °C in an Ar atmosphere (40 mL/min) for 10 min, reduced in $H_2$ (50 mL/min) for 5 min, and purged in Ar (40 mL/min) for 5 min to remove $H_2$ which hinders the formation of carbon networks on Cu surface. The growth step was soon followed at the same temperature in a diluted $C_2H_2$ atmosphere ($C_2H_2$:Ar = 1:100 mL/min) for 3 min. During this step, 50 W of ICP was generated to facilitate the dissociation of $C_2H_2$. The pressure was maintained at 50 mTorr during the whole process.

**Preparation of exfoliated samples.** As a reference, 1L samples supported on thermally grown $SiO_2$ ($G_{SiO2}$) were prepared by the mechanical exfoliation of natural graphite onto Si wafers topped with 285-nm $SiO_2$ layers. For $_{CuFilm}G$ samples, Cu films of 5 or 50 nm in thickness were deposited on top of $G_{SiO2}$ samples using a thermal evaporator.

**Optical and AFM characterizations.** The samples were characterized by Raman spectroscopy, optical microscopy, and atomic force microscopy (AFM). The details of the employed micro-Raman setup were given elsewhere.[36] Briefly, Raman spectra were obtained using three excitation photon energies of 2.71, 2.41, and 1.96 eV. The Raman signals back-scattered off the focal spot of ~1 μm in diameter were collected with an objective lens (40X, numerical aperture = 0.60) and guided into a spectrometer equipped with a charge-coupled device. The spectral resolution defined by the FWHM of the Rayleigh peak was 3 cm$^{-1}$, and the spectral accuracy was better than 0.5 cm$^{-1}$ for 1.96 eV. For polarized measurements, each polarization selected with an analyzing polarizer in front of the detector was scrambled by a quarter-wave plate to avoid the issue of polarization-sensitivity of the detector unit. The laser power was maintained as low as possible to avoid unwanted photoinduced effects. The topographic height images of samples were obtained in a non-contact mode using an AFM (Park Systems, XE-70).

## 3. Results and Discussion

In Fig. 1, we show the emission/scattering spectra of as-grown graphene on Cu foils ($G_{CuFoil}$) induced by photoexcitation at three different energies of 2.71, 2.41, and 1.96 eV. The spectra were dominated by the broad photoluminescence (PL) centered at 575 nm and also showed sharp Raman peaks. The PL signal originates from the interband transition between the s-p band crossing the Fermi level and d band located 2 eV below the Fermi level.[28] The spectrum obtained with 514 nm excitation revealed the G and 2D Raman peaks characteristic of single layer graphene[18-20] on top of a strong PL background. Whereas the D peak activated by structural defects[21] was also observed with noticeable spatial inhomogeneity as will be shown below, its intensity was much less than those of G and 2D peaks. The overwhelmingly strong PL baseline that degrades the signal-to-noise (S/N) ratio of the Raman peaks is a potential but obvious obstacle to efficient on-site Raman characterization of graphene grown on Cu substrates. Whereas Raman signals can be selected preferentially by various time gating methods in case the radiative decay is relatively slow,[37] the measurements require specialized instruments that may not be adequate for industrial settings for mass production. Instead, we varied excitation energy to minimize the overlap between the Raman and PL signals. As shown in Fig. 1, excitation with 2.71 eV resulted in significantly reduced PL background compared to the case using 2.41 eV that places G and 2D peaks near the center of the PL. Because of the $\nu^4$-dependence of Raman intensity,[38] 2.71 eV excitation gave a good S/N ratio despite the non-negligible PL background. The spectra obtained with 1.96 eV also showed reduced PL signals and exhibited larger D/G intensity ratios compared to the cases with higher photon energies. The latter fact consistent with a previous study[23] can be useful in quantifying disorder in as-grown graphene samples. Moreover, lower photon energy was advantageous because photoinduced oxidation of Cu substrates was less severe as will be explained below. The polarized detection also shown in Fig. 1 can be useful in suppressing the PL background. Whereas the G peak intensity was identical for the parallel and cross configurations of polarization,[39] the PL intensity was much weaker for the cross configuration, which became more evident with increasing photon energy: the cross/parallel intensity ratio decreased from

70% for 1.96 eV to 40% for 2.71 eV.

In Fig. 2a~2c, we show multiple representative Raman spectra obtained at 1.96 eV from three different types of samples: $G_{CuFoil}$, graphene/Cu film/SiO$_2$/Si ($G_{CuFilm}$), and Cu film/graphene/SiO$_2$/Si ($_{CuFilm}$G). Optical micrographs and AFM height images for each representative sample were given in Fig. S1. Whereas all three types of samples exhibited significant PL background, its intensity varied significantly among samples. In particular, the PL signals from $_{CuFilm}$G with 50 nm Cu film (Fig. 2c) were even stronger than others. In contrast, $_{CuFilm}$G with 5 nm Cu film showed negligible PL contribution as shown in Fig. S2. Figure S1c revealed that the thin Cu films primarily consisted of nanometer-scale islands and thus served as a quasi-dielectric medium. Transition to metal is known to occur at a thickness higher than 8 nm.[40] Non-negligible D intensity and significant spatial inhomogeneity for $G_{CuFoil}$ and $G_{CuFilm}$ indicate that the crystallinity of CVD-grown samples is worse than that of exfoliated samples ($_{CuFilm}$G). The spatial inhomogeneity is also known to cause the much broader line shapes for the CVD-grown samples,[41] as can be more clearly seen in Fig. 2d where representative Raman spectra of each type of sample were shown with polynomial-fit PL backgrounds subtracted. The G or 2D peaks of CVD-grown graphene exhibited significant upshifts when compared to pristine exfoliated graphene ($G_{SiO2}$). Interestingly, $G_{CuFoil}$ exhibited hardening in both of G and 2D peaks whereas $G_{CuFilm}$ did only in 2D peak. It is also notable that deposition of 50 nm Cu film on $G_{SiO2}$ led to downshift in both Raman peaks of $_{CuFilm}$G.

In order to confirm the statistical validity of the above spectral features, the frequencies of G and 2D peaks ($\omega_G$ and $\omega_{2D}$) obtained from multiple measurements of the above samples were projected in the $\omega_G$-$\omega_{2D}$ plot of Fig. 3. Notably, $G_{CuFoil}$, $G_{CuFilm}$, and $_{CuFilm}$G are grouped separately in ($\omega_G$, $\omega_{2D}$) space, which is consistent with Fig. 2d. It is well known that $\omega_G$ and $\omega_{2D}$ of graphene are sensitive to charge density[24, 25, 42] and lattice strain.[23, 24] Because of their distinctive effects on lattice vibrations, the two quantities can be precisely determined using the $\omega_G$-$\omega_{2D}$ plot.[29] The origin at (1581.6, 2629.3) cm$^{-1}$ that is denoted by O in Fig. 3 represents ($\omega_G$, $\omega_{2D}$) obtained at 1.96 eV[43] from freestanding graphene that is approximately charge-neutral and unstrained.[44] The red (black) dashed line depicts the trajectory that unperturbed graphene samples would follow upon the perturbation of hole doping (lattice strain). Thus, the fact that all the Raman map data of $G_{SiO2}$ (black squares) lie on the black line indicates that the whole area of $G_{SiO2}$ is nearly charge-neutral but with some spread in lattice strain. The slope for the strain axis in black was set as 2.6 to represent biaxial strain instead of 2.2 for uniaxial strain.[29] The slope of 2.6 for biaxial strain, an average of three available in the literature,[45-47] is consistent with the value that O. Frank et al. reported for their graphene grown on Cu.[33] Figure 3 showed that the data points for $_{CuFilm}$G with 5 nm Cu film (red squares) were hardly affected but those for 50 nm Cu film were displaced toward lower frequencies for both peaks when referenced to their pristine counterparts. Notably, however, all data points remained on the strain axis, which indicates that deposition of Cu films do not affect charge density within ~1x10$^{12}$ cm$^{-2}$ but imposes tensile stress on the underlying graphene when their thickness is sufficient. The deposition-induced change in the strain was ~0.3% for the 50 nm case but negligible for the 5 nm case. The observed stretch is attributed to adlayer-induced deformation of graphene that is partly suspended hills of underlying SiO$_2$ substrates.[48] Negligible deformation for the 5 nm case is also consistent with the fact that the percolation threshold for Cu film is thicker than 8 nm.

Unlike $_{CuFilm}$G, however, G$_{CuFoil}$ and G$_{CuFilm}$ were found to be located in the region (above the strain axis), which cannot be reached by mechanical deformation even in addition to charge injection.[29] It is also interesting that the displacement between the two groups is mostly parallel to the strain axis. We decompose the displacement of G$_{CuFoil}$ from the origin into two contributions, one originating from biaxial lattice strain (mechanical displacement) and the other from modification in the electronic structure of graphene (electronic displacement). Whereas the former leads to displacement along the strain axis, the latter affects only $\omega_{2D}$ and thus results in displacement along the blue dashed axis representing modification in the Fermi velocity ($v_F$) as will be explained below. The mechanical displacement of G$_{CuFoil}$ is induced by differential thermal expansion between graphene and Cu foil. Because the graphene has a negative thermal expansion coefficient unlike Cu,[49] CVD-grown graphene tends to expand against Cu while it cools from 1000 degree down to room temperature. However, strong adhesion[50] between graphene and Cu leads to in-plane compression of graphene that is forced by the contracting Cu foils. Quantitative analysis in Fig. 3 using experimental strain sensitivity[45-47] of $\omega_{2D}$ revealed that the CVD-induced compression amounts to 0.75 ± 0.08%. Relaxation of such compressive strain was confirmed for CVD-grown graphene when transferred onto other substrates via wet etching of Cu foils.[51]

Notably, that G$_{CuFilm}$ exhibited minimal mechanical displacement (Fig. 3), which amounts to a tensile strain of 0.1 ± 0.05%. The reduction can be attributed to the fact that the Cu film of G$_{CuFilm}$ is only 300 nm thick and attached to thermally grown 300-nm thick SiO$_2$ layer that has two orders of magnitude smaller thermal expansion coefficient than Cu.[52] Backing Cu catalysts with silica or other materials with small thermal expansion can be useful in growing graphene with minimal built-in mechanical strain. We note that the vertical electronic displacement shown in Fig. 3 was also found in graphene grown by Cu-CVD by others.[34, 53] Solid squares in Fig. 3 denote average $\omega_G$ and $\omega_{2D}$ of graphene grown on Cu foils and single crystals of three different facets by Frank et al.[33] Despite the dependence of strain on the crystalline facets, all their values of ($\omega_G$, $\omega_{2D}$) are located within the forbidden region. It is to be noted that their samples grown on Cu foils showed much less compressive strain than ours despite identical nominal growth temperature, which implies that the built-in strain is also affected by processing parameters other than temperature.

Whereas the $\omega_G$-$\omega_{2D}$ plot[29] has been widely used in quantifying strain and charge density, it is approximate metrology based on the assumption that the two variables are the only factors affecting the Raman frequencies and independent of each other. In the double resonance scattering process,[32] 2D phonons of higher frequency are generated[31] as the Fermi velocity ($v_F$) is decreased because of van der Waals interaction with environments.[30] In contrast, G peak originating from the zone-center phonon is not shifted because such a modest electronic perturbation hardly affects the energy of the G phonon. Thus caution must be paid in using the $\omega_G$-$\omega_{2D}$ plot for the cases where interaction with substrates modifies the electronic structure of graphene. As shown for graphene supported on hexagonal BN crystals, $\omega_G$ and $\omega_{2D}$ indeed depended not only on the strain and charge density but also on effective $v_F$.[41] In principle, unambiguous decomposition of a given ($\omega_G$, $\omega_{2D}$) into the three variables is not possible without additional information regarding them.[41] In order to fix one of the variables, we assumed that charge doping by Cu substrates is negligibly small, which is supported by our measurements with $_{CuFilm}$G (Fig. 3). Although some reported that CVD-grown graphene is doped with electrons, there are also experimental observations that the doping is negligibly small[34] and thermally activated.[54] Then, the displacement of each ($\omega_G$, $\omega_{2D}$) from O can be decomposed along the two axes for strain and $v_F$. Assuming the linear dispersion of the π bands, the fractional change in the effective $v_F$ is proportional to the change in 2D frequency (d$\omega_{2D}$) as follows:[31]

$$\frac{dv_F}{v_F} = -\frac{\hbar v_F}{\left[E_L - \frac{\hbar \omega_{2D}}{2}\right]\frac{d\omega_{2D}}{dq}} d\omega_{2D} \quad \text{(Equation 1)}$$

, where $E_L$ and q are excitation photon energy, and wave vector of phonon, respectively. The vertical axis shown as a blue dashed line in Fig. 3 represents the percentile reduction in $v_F$ estimated for $E_L$ = 1.96 eV using that $\hbar v_F = 6.5$ eV$\dot{A}$, $\omega_{2D} = 2629$ $cm^{-1}$,[43] and $\frac{d\omega_{2D}}{dq} = 0.08$ $eV\dot{A}$.[31] Figure 3 readily revealed that the interaction with Cu catalysts decreased $v_F$ by 9.5 ± 3 and 14.6 ± 2 % for $G_{CuFoil}$ and $G_{CuFilm}$, respectively. It is to be noted that the reduction is referenced to graphene supported on thermally grown $SiO_2$.[29]

We now show that the reduction in $v_F$ is strongly dependent on the excitation energy. For the sake of statistics, multiple spots from several $G_{CuFoil}$ samples were probed with three different excitation lines, and their data are given in Fig. 4 with average values. Because most sample areas looked similar without conspicuous landmarks under an optical microscope (Fig. S1a), it was not always possible to aim at an identical spot with different excitation lasers. Despite the potential variance in target spots, however, $\omega_G$ remained within a narrow range regardless of their excitation energy. In contrast, $\omega_{2D}$ varied significantly because of the photon energy-dependent dispersion.[55, 56] For vector decomposition of strain and $v_F$ as done in Fig. 3, the origin, strain axis, and percentile reduction in $v_F$ were given for each excitation energy[43] in Fig. 4. The strain in the selected samples amounted to 0.75 ~ 0.85 % on average, and the slight differences for different excitation energies can be attributed to spatial inhomogeneity. The vertical displacement of $\omega_{2D}$ off the strain axis ($\Delta\omega_{2D}$) was 17 ± 5.8 cm$^{-1}$ for 1.96 eV and decreased to 2.4 ± 6.8 cm$^{-1}$ for 2.71 eV as also shown in the inset of Fig. 4. Because the magnitude of $\Delta\omega_{2D}$ is approximately proportional to the photon energy according to the above equation, the opposite change in $\Delta\omega_{2D}$ suggests that the effective $v_F$ is strongly dependent on the photon energy. The percentile reduction in $v_F$ (inset of Fig. 4) indeed shows that the effective $v_F$ of $G_{CuFoil}$ is smaller for higher excitation energy.

The unusual upshift of $\omega_{2D}$ that cannot be attributed to strain or charge doping has been observed for graphene in contact with another graphene,[31] hexagonal BN,[41] and metallic catalysts for CVD.[33, 34, 53] Whereas the shift was attributed to a decrease in $v_F$ for some cases,[31, 34, 41] it was also ascribed to modified phonon dispersion[57, 58] or competition between the inner and outer processes in the double resonance.[53] We note that the nonlinear electronic dispersion of graphene must be considered to resolve the controversy. In the single-particle picture where inter-electronic (e-e) Coulomb interactions are neglected, the theory predicts that the π bands of graphene (blue lines in Fig. 5) have a linear dispersion with a slope of $\hbar v_F$.[59] In the presence of the interactions, however, theoretical divergences are encountered as the density of states approaches zero at the charge neutrality point.[60] Consequently, the Dirac cones become deformed in such a way that the slope diverges near the neutrality point (red line in Fig. 5).[59] Indeed an extremely high $v_F$ of $3\times10^6$ m/s was observed for freestanding graphene devices.[59] Hwang et al. also showed that $v_F$ could be varied in the range of $1.15 \sim 2.49\times10^6$ m/s by using substrates of different dielectric constants.[61]

Such changes in the electronic dispersion will lead to shifts in $\omega_{2D}$ as will be explained below using the scheme for DR process for 2D peak (Fig. 5).[32, 62] The blue linear lines denote the π bands of $G_{CuFoil}$ that were approximated to have a linear dispersion because of the high dielectric constant of Cu. The red curvy lines represent $G_{SiO2}$ that served as a reference system for the $\omega_G$-$\omega_{2D}$ metrology. Whereas early measurements using Shubnikov de Haas oscillations resulted in $v_F$ of $\sim1.05\times10^6$ m/s for $G_{SiO2}$ under the assumption of a linear dispersion,[63, 64] a more recent angle-resolved photoemission (ARPES) study showed that the π bands are noticeably bent for graphene supported on dielectric substrates of hexagonal BN and quartz.[61] During the DR process for $G_{CuFoil}$, the π-π* excitation will be accompanied by the generation of two D phonons that have larger wave vector ($q_{Cu}$) than those ($q_{SiO2}$) of D phonons generated for $G_{SiO2}$. This fact explains why $\omega_{2D}$ is unusually higher for $G_{CuFoil}$ than $G_{SiO2}$. Then the percentile reduction in $v_F$ extracted from Equation 1 corresponds to the slope difference between the blue line and the dashed line in Fig. 5a. The photon energy dependence of $\Delta\omega_{2D}$ shown in Fig. 4 can also be explained by the fact that the slope difference should decrease for higher excitation energy, as depicted in Fig. 5b.

The experimental electronic structures of graphene mostly agree with our results. J. Avila et al. reported that $G_{CuFoil}$ has a linear dispersion with $v_F$ of $\sim1.0\times10^6$ m/s.[65] C. Hwang et al. determined the curved π bands for graphene samples grown on SiC, supported on hexagonal BN, and supported on quartz.[61] It is evident that graphene on dielectric substrates undergoes reduced screening and has a steeper slope than $G_{CuFoil}$. Although these previous results are consistent with upshifts of $\omega_{2D}$ for $G_{CuFoil}$ with respect to $G_{SiO2}$, however, they cannot be based on to validate the photon energy dependence of $\Delta\omega_{2D}$ (Fig. 4) quantitatively because of lacking information for $G_{SiO2}$. Whereas quartz is very close in stoichiometry and dielectric constant to thermal $SiO_2$, it was shown that the ARPES results from the two types of substrates[61, 66] were significantly different, which was attributed to the presence of interfacial impurities.[61] Besides, the early ARPES-derived π bands for $G_{SiO2}$ lack accuracy because of the poor signal-to-noise ratio.[61] Further validation will require better measurements of the π bands of $G_{SiO2}$.

Our results clearly show that the degree of adhesion between graphene and Cu substantially affects $\Delta\omega_{2D}$ and thus the effective dielectric environment experienced by graphene. Thermal deposition of Cu film on top of $G_{SiO2}$ did not induce electronic displacement in $\Delta\omega_{2D}$ (Fig. 3). As seen for $_{CuFilm}G$ with 5 nm Cu film, copper does not wet graphene well and forms nanometer-scale islands with numerous voids at the graphene-Cu interface. In such a geometry, dielectric screening of electrons in graphene is expected to be inefficient despite nominal physical contact with Cu films. Similar observations were made for graphene wet-transferred onto Cu foils.[34] Strong coupling between graphene and Cu foils in $G_{CuFoil}$ could also be broken by photooxidation of Cu substrates. As shown in Fig. S3, the surface of Cu foils remained intact during repeated 4 measurements with 2.71 eV excitation laser (Fig. S3a) but underwent oxidation for prolonged irradiation (Fig. S3b). It is to be noted that 2D peak downshifted as Raman peaks for $Cu_2O$ emerged. Figure S3c depicting $(\omega_G, \omega_{2D})$ extracted from Fig. S3b showed that the unoxidized state (i) is electronically displaced in $\omega_{2D}$. The trajectory of $(\omega_G, \omega_{2D})$ further showed that the photoinduced interfacial oxides essentially decouple graphene electronically from Cu substrates (ii) and impose tensile stress (iii & iv). The change in the lattice strain between i and iv was ~0.4%.

## 4. Conclusions

In this work, we performed multi-wavelength Raman spectroscopy for graphene in contact with various forms of Cu to establish Raman metrology for graphene grown by Cu-CVD. To improve the signal-to-noise ratio of Raman spectra, overwhelmingly large photoluminescence signals of Cu substrates could be partially avoided by using high (or low) excitation photon energy and polarized detection. The large upshifts of G and 2D peaks of as-grown graphene were mainly attributed to thermally induced lattice strain and could be significantly attenuated by backing Cu catalyst films with $SiO_2$ substrates with low thermal expansion coefficient. The significant fraction of 2D upshifts that could not be attributed to strain nor charge doping turned out to originate from the electronic coupling with underlying Cu, which modifies the electronic bands and thus affects the double resonance scattering of 2D peak. We also showed that the electronic displacement of 2D peak requires intimate physical contact, which thermally deposited Cu films on graphene lacked. This finding indicated that it is the dielectric environment in close proximity of graphene that influences its Raman scattering. Overall, the Raman analysis presented in the current study will serve as an efficient optical method that allows simultaneous quantification of lattice strain and electronic coupling in graphene directly bound on solid substrates.

**Conflicts of interest:** The authors declare no conflict of interest.

**Acknowledgements**

This work was supported by the National Research Foundation of Korea (No. 2016R1A2B3010390 and No. 2016M3D1A1900035).

**Appendix A. Supplementary data**

Supplementary data related to this article can be found elsewhere.

**Figures and Captions**

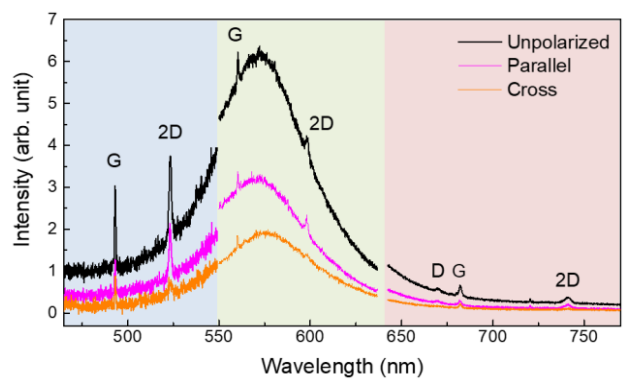

**Figure 1. Multi-wavelength Raman scattering/photoluminescence (PL) spectra of graphene grown on Cu foil (G$_{CuFoil}$).** Spectra in each color-shade were excited at 457 nm (2.71 eV in blue), 514 nm (2.41 eV in green), and 633 nm (1.96 eV in red shade). Each exhibited D, G or 2D Raman peaks from graphene on top of strong PL background from Cu. Magenta and orange curves were obtained in a parallel and cross-polarization configuration, whereas the black curve was obtained without an analyzer.

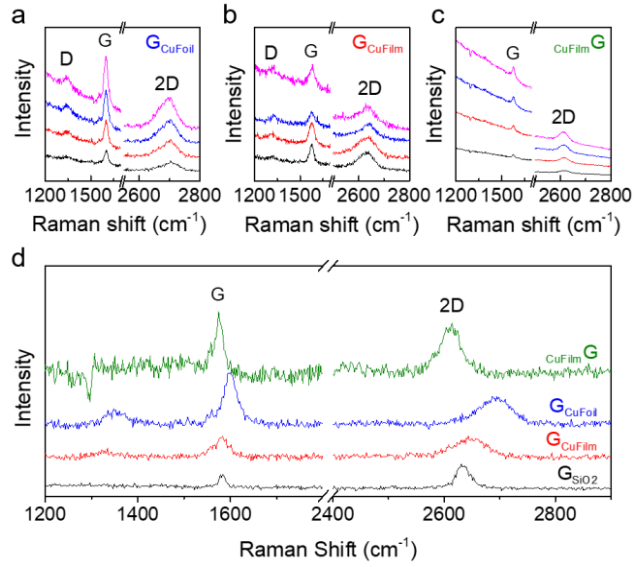

**Figure 2. Raman spectra of graphene bound on Cu in various forms.** (a) $G_{CuFoil}$, (b) graphene on Cu film/SiO$_2$/Si substrates ($G_{CuFilm}$), (c) graphene deposited with Cu film of 50 nm in thickness ($_{CuFilm}G$). Each panel presented four selected spectra obtained at 1.96 eV to show typical sample-to-sample variation. (d) Representative spectra for each type of sample shown with PL from Cu subtracted. Raman spectrum of graphene on SiO$_2$/Si substrate ($G_{SiO2}$) was shown in black for comparison.

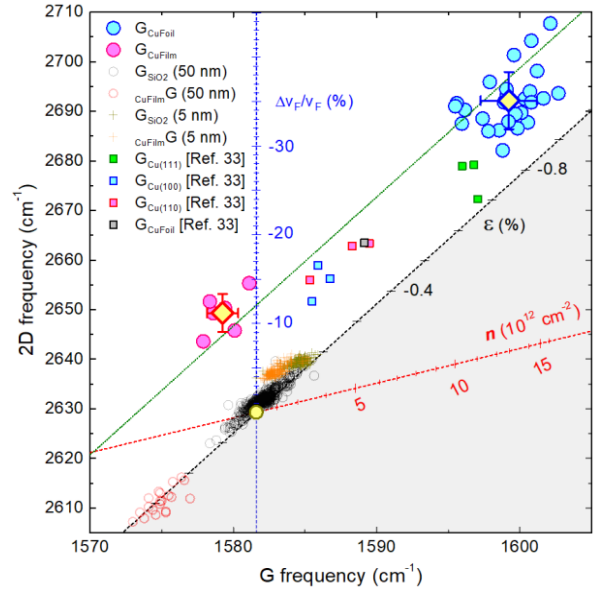

**Figure 3. Distinctive displacements of G and 2D frequencies of $G_{CuFoil}$ and $_{CuFilm}G$.** Each point represents one pair of G and 2D frequencies ($\omega_G$, $\omega_{2D}$) obtained from ~1 μm² spot of $G_{CuFoil}$ (blue circles) and $G_{Cufilm}$ (red circles). Yellow-filled diamonds represent statistical averages with standard deviations in error bars. Data for $_{CuFilm}G$ were obtained before and after the deposition of 5 and 50 nm Cu film on top of $G_{SiO2}$. For comparison, data of graphene grown on Cu foil and single crystals of three facets were added (Ref. 33). For the details of the origin (yellow circle) and three dashed lines for charge density ($n$), strain ($\varepsilon$), and fractional Fermi velocity reduction ($\Delta v_F/v_F$), see the main text. The green dotted line passes through the two yellow-filled diamonds in parallel with the black dashed line.

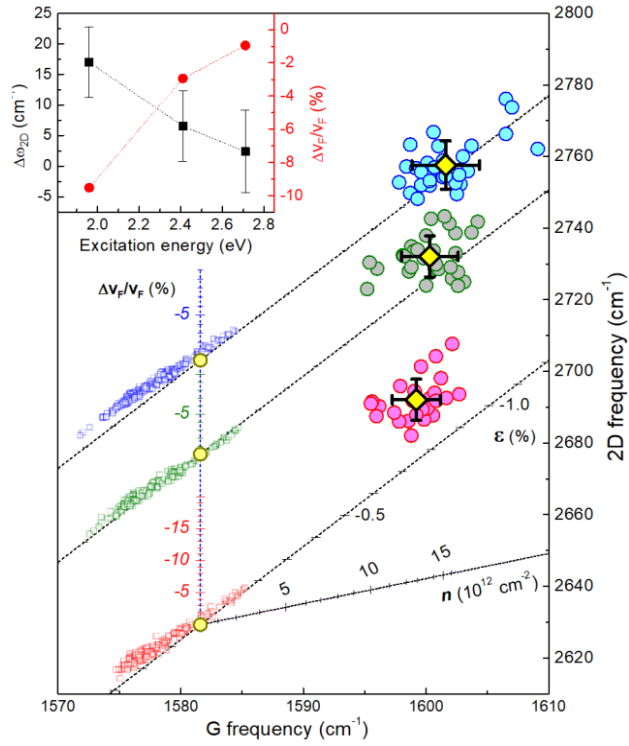

**Figure 4. Photon-energy dependence of 2D frequency shift of $G_{CuFoil}$.** Multiple ($\omega_G$, $\omega_{2D}$) data were obtained for $G_{CuFoil}$ at three energies of 2.71 (blue), 2.41 (green), and 1.96 eV (red). Sets of squares near the origins represent $G_{SiO2}$ (adopted from Ref. 43). Inset shows electronic displacement of 2D peak (left ordinate) and fractional reduction in Fermi velocity (right ordinate).

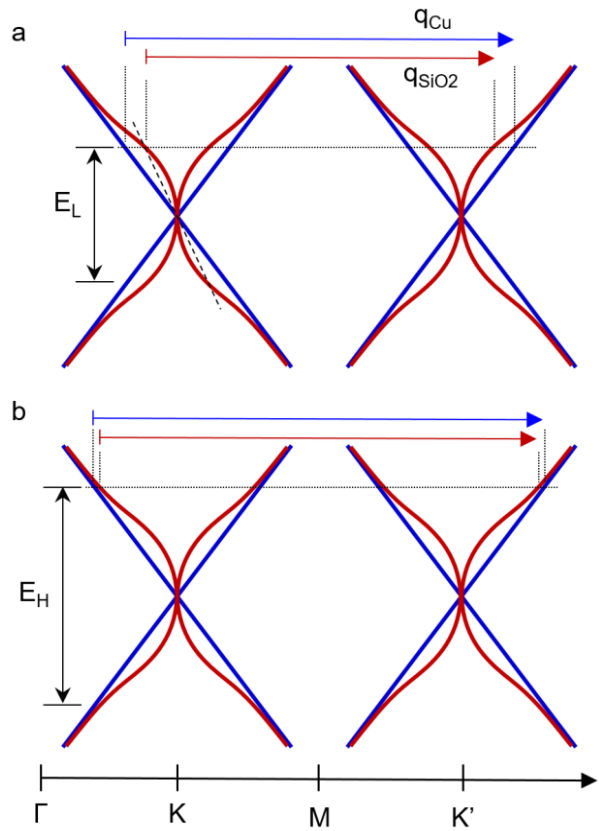

**Figure 5. Scheme for electronic displacement of 2D peak.** Double resonance (DR) scattering process for 2D peak at low (a) and high (b) excitation energies ($E_L$ and $E_H$). Electronic bands of graphene across high symmetry points in Brillouin zone are represented by blue lines for $G_{CuFoil}$ and red lines for $G_{SiO2}$ (see the main text for linear and nonlinear dispersion). Momenta of D phonons selected during DR process are denoted by blue ($G_{CuFoil}$) and red ($G_{SiO2}$) arrows.